\documentclass{aa}

\usepackage{subcaption}
\usepackage{graphicx}
\usepackage{txfonts}
\usepackage[breaklinks]{hyperref} 
\begin{document}

 \title{Alfvénic waves in polar spicules}


\author{E. Tavabi\inst{1,2}
         \and
          S. Koutchmy \inst{1}
         ,
          A. Ajabshirizadeh \inst{3,4}
         ,
          A. R. Ahangarzadeh Maralani \inst{3,4}
         \and
          S. Zeighami \inst{3,4}
                 }

\institute{Institut d’Astrophysique de Paris, UMR 7095, CNRS and UPMC, 98 Bis Bd. Arago, 75014 Paris, France \\
              \email{koutchmy@iap.fr}
         \and
             Physics Department, Payame Noor University (PNU), 19395-3697-Tehran, I. R. of Iran\\
             \email{tavabi@iap.com}
         \and
             Center for Excellence in Astronomy \& Astrophysics (CEAA), Research Institute for Astronomy \& Astrophysics of Maragha (RIAAM), Maragha, Iran, P.O.Box: 55134-441\\
             \email{ali\_ajabshir@yahoo.com}
         \and
             Department of Physics, Tabriz Branch, Islamic Azad University, Tabriz, Iran\\
             \email{ahangarzadeh@iaut.ac.ir}\\
             \email{zeighami@iaut.ac.ir}\\
          }



  \abstract
   {For investigating spicules from the photosphere to coronal heights, the new Hinode/SOT long series of high resolution observations from Space taken in CaII H line emission offers an improved way to look at their remarkable dynamical behavior using images free of seeing effects. They should be put in the context of the huge amount of already accumulated material from ground-based instruments, including high- resolution spectra of off-limb spicules.}
   {Both the origin of the phenomenon and the significance of dynamical spicules for the heating above the top of the photosphere and the fuelling of the chromospheric and the transition region  need more investigation, including a possible role of the associated magnetic waves for the corona higher up. }
   {We analyze in great detail the proper transverse motions of mature and tall polar region spicules for different heights, assuming that there might be Helical-Kink waves or Alfvénic waves propagating inside their multi-component substructure, by interpreting the quasi-coherent behavior of all visible components presumably confined by a surrounding magnetic envelop. We concentrate the analysis on the taller CaII spicules more relevant for coronal heights and easier to measure. \\
   2D velocity maps of proper motion were computed for the first time using a correlation tracking technique based on FFTs and cross-correlation function with a 2nd-order accuracy Taylor expansion.  Highly processed images with the popular mad-max algorithm were first prepared to perform this analysis. The locations of the peak of the cross-correlation function were obtained with sub-pixel accuracy.}
   {The surge-like behavior of solar polar region spicules supports the untwisting multi-component interpretation of spicules exhibiting helical dynamics. Several tall spicules are found with (i) upward and downward flows similar at lower and middle-levels, the rate of upward motion being slightly higher at high levels; (ii) the left and right-hand velocities are also increasing with height;  (iii) a large number of multi-component spicules show shearing motion of both left-handed and right-handed senses occurring simultaneously,which might be understood as twisting (or untwisting) threads. The number of turns depends on the overall diameter of the structure made of components and changes from at least one turn for the smallest structure to at most two or three turns for surge-like broad structures; the curvature along the spicule corresponds to a low turn number similar to a transverse kink mode oscillation along the threads.}
   {}
   \maketitle
   \section{Introduction}
     Spicules are jet-like chromospheric structures and are usually seen all around the limb of the Sun, see \cite{Tsiropoula} for a recent exhaustive review presentation with a rather conservative point of view and covers well the ground-based observations. As far as time sequences of images are concerned, ground-based observations of extremely fine off-limb spicules have suffered from the terrestrial turbulence effects. Spicules arise in different directions in the low-level interface between the photosphere and the corona. Their emission in low excitation emission lines is responsible for the part of the interface usually called the chromospheric shell where HI, HeI and HeII high FIP (First Ionization Potential) emission lines are seen and further out, towards the higher temperature transition region (TR) (see \citealt{Bazin} for eclipse observations free of parasitic scattered light). Indeed, eclipse observations showed for a long time that at the lowest 4 Mm heights the temperature remains low \citep*{Matsuno}. The mechanism of spicule formation and their evolution is not well understood  (see the reviews by \citealt{Sterlingc};\citealt{Zaqarashvili} and by \citealt{Mathioudakis}; for different propulsive mechanisms, see e.g. \citealt{Lorraina} and \citealt{Lorrainb};\citealt{Filippovb} and \citealt{Tavabia}). The investigation of solar spicules is necessary to understand the Transition Region and the chromosphere and possibly some aspect of coronal heating (\citealp{Kudoh}) especially in the case of tall or giant spicules that have more chance to contribute in the corona. The mature spicules are rather homogeneous in height and have a lifetime of approximately 5-15 min which is comparable to the photospheric granules lifetime. They have typical upflow speeds of 20-50 kms$^{-1}$ and sometimes much higher for giant spicules. Spicule diameters in the chromosphere are of the order of 200-500 km when individual components of shorter apparent lifetime are considered that are now often called type II spicules, starting with \cite{De Pontieub}. Some objections regarding their significance were expressed (\citealp{Zhang};\citealp{Klimchuk});  a more detailed and documented description was also given based on excellent ground-based SST spectroscopic observations and analysis (\citealp{De Pontieuc}) that we found in agreement with much earlier results by \cite{Dara}. Indeed, type II spicules were also discussed in the frame of macro-spicules and even “blow-out SXR coronal jets” by \cite{Sterlinga} and \cite{Sterlingb}, confirming they are the components of a more significant event.\\

     Following the results of our statistical and morphological analysis of spicules (\citealp{Tavabia};\citealp{Tavabid} and \citealp{Tavabie}), we further will not really distinguish type II spicules from the larger structure they are a part of. Note that the corresponding unseen magnetic flux tube or envelop or shell surrounding a set of spicule components is evidently larger than the spicule itself. They usually reach heights of typically 4 Mm and up to 10 Mm before fading out of view or fall free back towards the solar surface. Their smallest widths can be only 100-200 km, close to the resolution limit of the best telescopes. \cite{Tavabia} found that indeed spicules show a whole range of diameters, including unresolved "interacting spicules" (I-S), depending on the definition chosen to characterize this ubiquitous dynamical phenomenon occurring inside a low TR and coronal background. Spicules are very thin and numerous, so along the line of sight some overlapping could occur, especially near and above the limb where a long integration along the line of sight exists when the projection in the plane of the sky is considered. Superposition effects (overlapping) are more important than was previously anticipated, when it was thought that spicules have at least 1 Mm or more diameter, because the number of spicules intercepted along the line of sight per resolution element is indeed considerable. A kind of collective and quasi-coherent behavior of 2 or more components of spicules was described for the first time in \cite{Tavabia} and \cite{Tavabi}. Most spicules were found to have a multiple structure (similarly to the doublet spicules described in \citealp{Suematsua} and \citealp{Tsiropoula}) and show impressive transverse periodic behavior which was interpreted as upwardly propagating kink or Alfvén waves.\\

    Spicules usually show an oscillatory behavior (\citealp{Zaqarashvili}). The existence of 5 minute oscillations in spicules were originally reported by \cite{Nikolsky} and \cite{Kulidzanishvili}, and others including spectroscopically performed observations with a large aperture coronagraph. More recently, image sequences were studied by \cite{De Pontieua}; \cite{Xia} and \cite{Ajabshirizadeh}. Oscillations in spicules with even shorter periods have been reported by \cite{Nikolsky}. They found that spicules oscillate along the limb with a characteristic period of about 1 min, a remarkable result confirmed by recent analysis of SOT (Hinode) spicules showing <120 sec  transverse oscillations from time-slice analysis (\citealp{Tavabia}). \cite{Kukhianidze} reported periodic spatial distribution of Doppler velocities with height through spectroscopic analysis of $H\alpha$ series in solar limb spicules (at the heights of 3800-8700 km above the photosphere).\\

    \cite{Suematsub} reported the observation of twist motion of spicules from Hinode SOT image observations and \cite{He} found high frequency transverse motion in spicules, also in agreement with the results of \cite{Tavabia}. Additional statistically significant parameters should be determined to permit a full interpretation of the phenomenon, including a possible spinning motion. Rotation about the quasi-radial axis of spicules has been inferred in several early classical spectroscopic studies (e.g. \citealp{Pasachoff}) from the tilts observed on slit spectra, without fully resolving the components of spicules (\citealp{Dara}) and the suggestion of spinning and/or helical motion in emerged flux tubes is a recurrent suggestion by theoreticians considering the currents inside (e.g. \citealp{Shibata}) and by observers (e.g. \citealp{Rompolt}).


\section{Observations}

   We selected two sequences of solar limb observations made in the Polar region with the broad-band filter instrument (BFI) of the Hinode SOT (Figure1). We used several series of image sequences obtained on 17 June 2011 (South Pole) and 25 October 2008 (North Pole) in the Ca II H emission line, the wavelength pass-band being centered at 398.86 nm with an FWHM of 0.3 nm . A fixed cadence of 1.6 s (for 25 Oct. 2008 it is about 10 s) is used (with an exposure time of 0.5 s) giving a spatial resolution of the SOT/Hinode observations limited by the diffraction pattern to 0.16 arcsec (120 km); a 0.1 arcsec pixel size scale is used. The images size for 17 June 2011 is 512$\times$512 pixels (readout performed only over the central pixels of the larger detector to keep the high cadence within the telemetry restrictions) that covers an area of (FOV) 55.78$\times$55.78 arcsec$^{2}$ and the images size is 512$\times$1024 and FOV about of 55.78$\times$111.57 arcsec$^{2}$ for 25 Oct. 2008 observations. On the polar cap of the Sun, spicules are somewhat more numerous than at low latitudes close to the solar equator, and they are slightly taller and definitely oriented more radially (\citealp{Filippova}). We used the SOT routine fg$\_$prep to reduce the image spikes and jitter and to align the time series. Note that the time series showed a slow instrumental drift, with an average speed smaller than 0.015 arcsec/min toward the north, as identified from solar limb motion.\\

\begin{figure}
   \centering
   \includegraphics[width=\hsize]{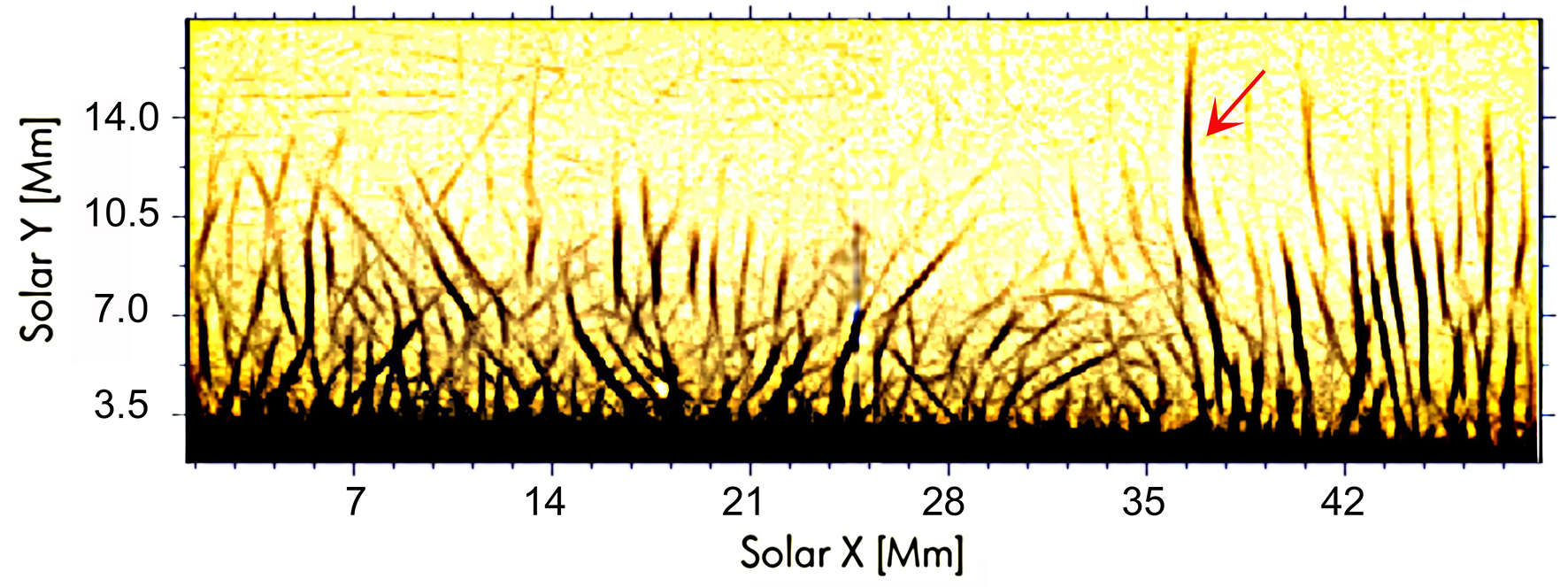}
      \caption{ Example of an image processed by the mad-max algorithm. The image is based on a Hinode/SOT narrowband Ca II H image taken on 17 June 2011 at 9:13:42.4 UT. The color table is inverted to produce a negative image. The arrow point to the long twisted spicule and the loop structures next to it. A movie showing the temporal evolution is available in the online edition (the spinning motions in the movie are most obvious when the movie is manually run at a higher speed than the default speed).
              }
   \label{figure1}
   \end{figure}

\begin{figure}
   \centering
   \includegraphics[width=\hsize]{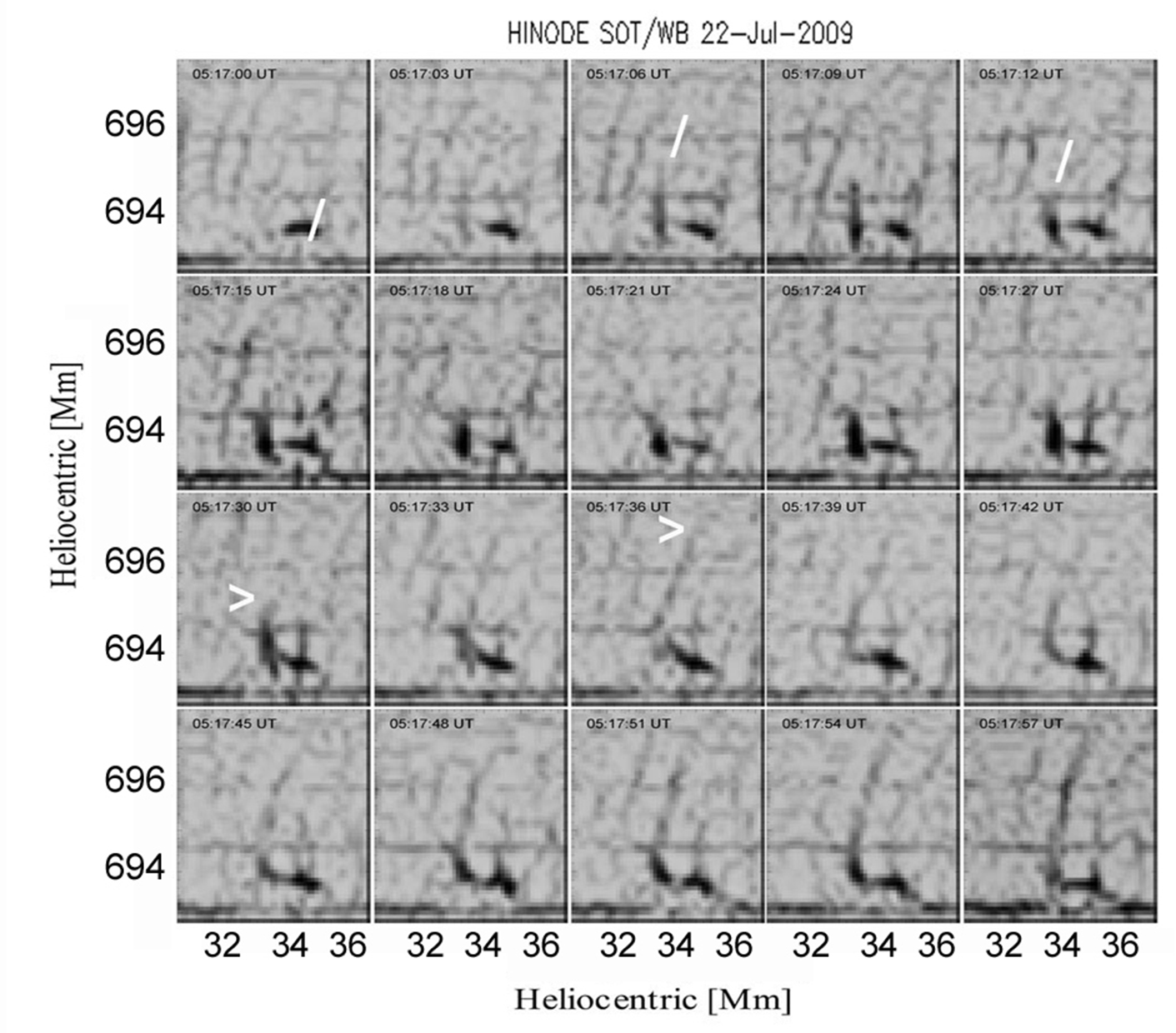}
      \caption{ Example of rapidly moving features observable at the ultimate SOT resolution, after using the mad-max processing and a negative image display of the results. Only the first 5 Mm heights above the limb are shown.
              }
         \label{figure2}
   \end{figure}

   The analysis is performed after the data are subjected to a deep processing for showing thread-like features; this is obtained using the mad-max algorithm (\citealp{Koutchmy};\citealp{Tavabid} and \citealp{Tavabie}) see Fig.~\ref{figure1} for a sample. This spatial non linear filtering clearly shows rather bright radial threads in the chromosphere as fine as the resolution limit of about 120 km, without any doubt confirming the high quality of SOT Hinode observations. It permits us to evaluate, in first order approximation, what could be the properties of individual spicules. Unfortunately, the signal/noise level does not permit us to utilize the full spatio-temporal resolution, as illustrated on Fig.~\ref{figure2}. Accordingly we will further restrict our analysis to the longest features, specifically the ones that can be automatically measured with a computer aided algorithm, leaving for a future paper the investigation of the ultimate very dynamical details seen in Fig.~\ref{figure2}.\\

\section{Data Analysis}

   Regarding the image processing, we found superior results after using a spatial image processing for both thread-like (or elongated) and loop like features obtained with the so-called mad-max algorithm (\citealp{Koutchmy}). The mad-max operator acts to substantially enhance the finest scale structure. The mad-max operator is a weakly nonlinear modification of a second derivative spatial filter which improves the resolution limit near the Nyquist frequency where the signal/noise is evidently critical. Specifically, it is where the second derivative has a maximum when looking along different directions (usually, 8 directions around each pixel are used to make the image processing fast). The behavior of mad-max qualitatively resembles the second derivative, but the strong selection effect for the direction of the maximum variation substantially enhances the intensity modulations of the most significant structures and, accordingly, considerably reduces the noisy background usually  appearing in high spatial filtering near the Nyquist frequency. It also appears to reduce the blending (due to overlapping effects) between crossing threads superposed along the line of sight. The algorithm, as originally proposed, samples the second derivative in eight directions, but the directional variation of the second derivative was generalized to a smooth function with a selectable passband spatial scale for this work (for more details see \citealp{Novembera} and the more recent paper by \citealp{Tavabie}). Spatial filtering using “mad-max” algorithms clearly shows relatively bright radial threads in the chromosphere as fine as the resolution limit of about 120 km (Fig.~\ref{figure1}). Note that some deviation from the radial direction is observed and the aspect ratio of each spicule is often larger than 10.\\
\section{Results}

\begin{figure}
   \centering
   \includegraphics[width=\hsize]{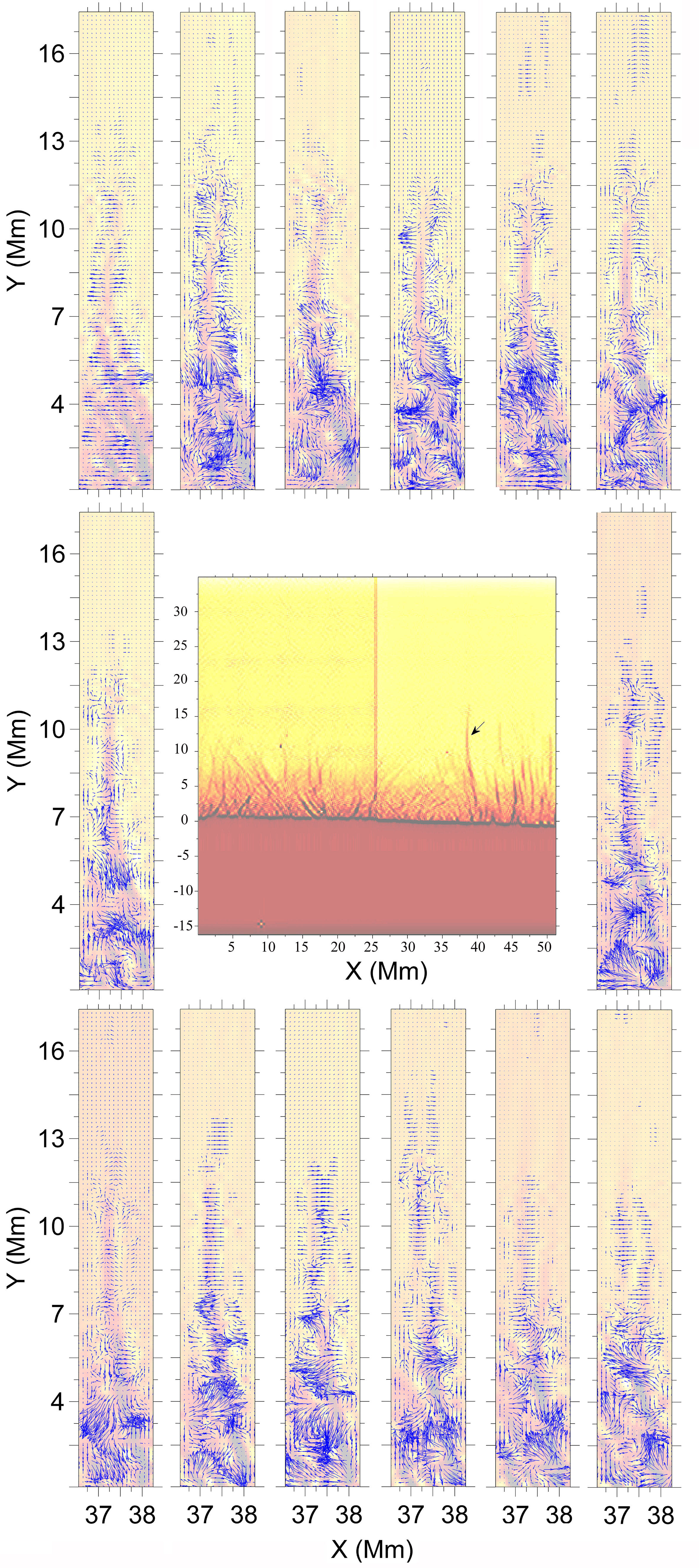}
      \caption{ Example obtained using the FLCT algorithm for showing the 2D velocity maps from successive frames with a corresponding remarkable structure shown with the red arrow over a larger field of view shown in negative at the center. Spatial units used in the display are 0.1 arcsec (corresponds to pixel size), and the cadence is 1.6 sec. Intensities in the proper motion maps are reproduced in red-orange color. Note the large magnification needed to clearly evaluate the results (see in better reproduction in the electronic color version, to be magnified), see Fig.~\ref{figure4} for a quantitative evaluation of the amplitudes of velocities.
              }
         \label{figure3}
   \end{figure}

   The selected spicules show a peculiar structure like as indicated in Fig.~\ref{figure3} at center. Only a small part of all spicules seen in the time series are shown, with a definite twisting effect and isolated enough to avoid superposition effects. Ten very long and obvious spicules in each time series were studied in details, after selecting the features where the overlapping effect is less important and where spicules are rather taller (see the example shown in Fig.~\ref{figure1}). We chose an axis close to the local vertical axis for each case, to follow the details with respect to this relative axis. Further, we used the Fourier Local Correlation Tracking algorithm (FLCT) to map the apparent proper motions over the field of view of each case. This is a powerful cross-correlation technique for measuring proper motions of tracers seen on successive frames of a time series of solar features. The original concept of the FLCT algorithm was published by \cite{Novemberb} and later the algorithm was improved by \cite{Fisher}. They described computational techniques to construct a 2D velocity field that connects two successive images taken at two different times; one must start from some given location within both images, compute a velocity vector, and then repeat the calculation while varying that location over all pixel positions. The cross-correlation is define as a function of position in the image, within a spatially localized apodization window by Gaussian function with a typical width. The distance between spatially localized cross-correlation maxima divided by the time interval, gives a measure of the proper motion projected in the plane of the sky.\\

\begin{figure}
  \centering
     \includegraphics[width=\hsize]{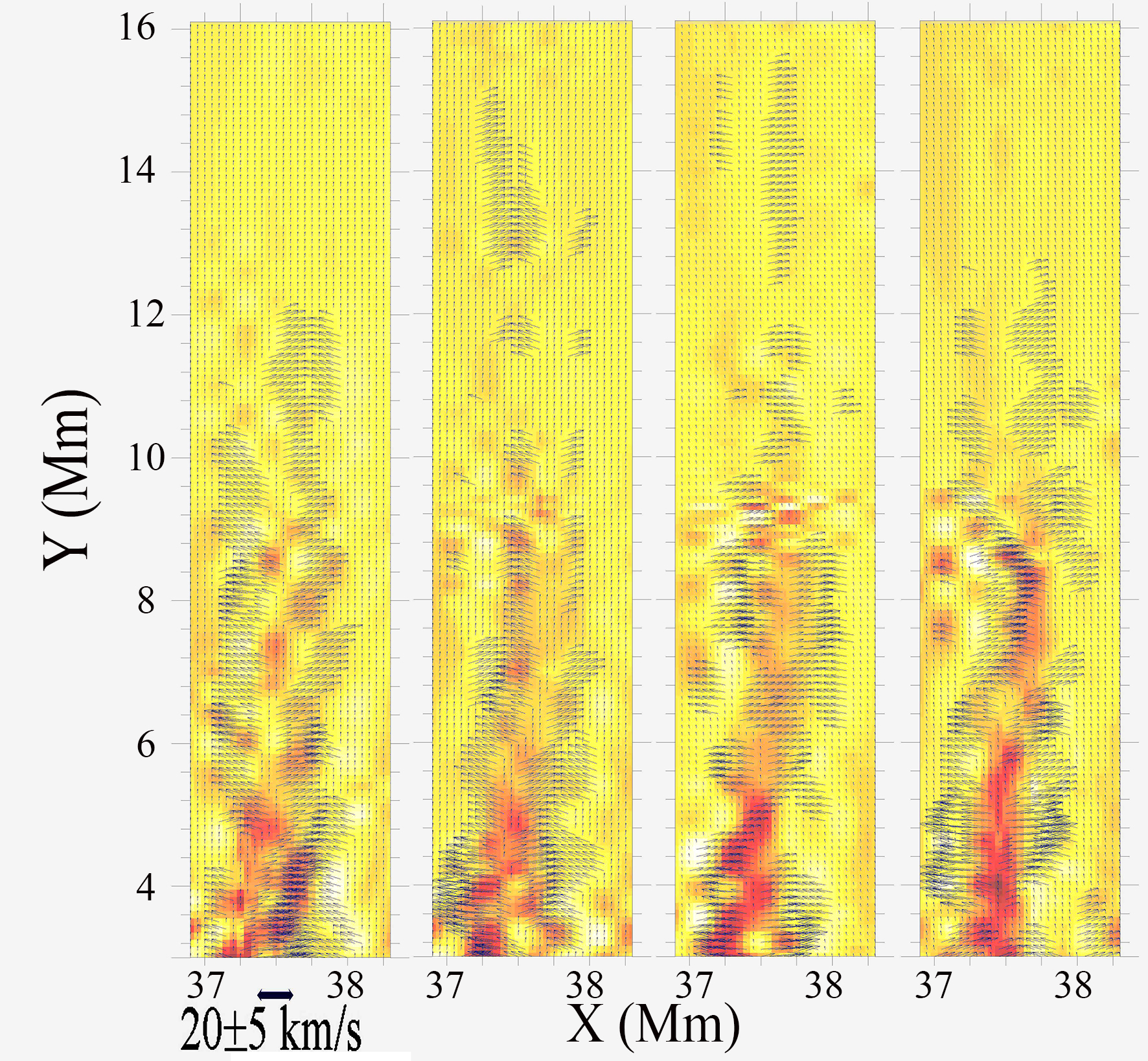}
      \caption{Selected frames from the left top first raw of the preceding Fig.~\ref{figure3}. The scale of apparent velocities is shown at the bottom using a small black bar put over the typical value affected by an error of $\pm$ 5 kms$^{-1}$.
              }
         \label{figure4}
\end{figure}

\begin{figure}
   \centering
   \includegraphics[width=4cm]{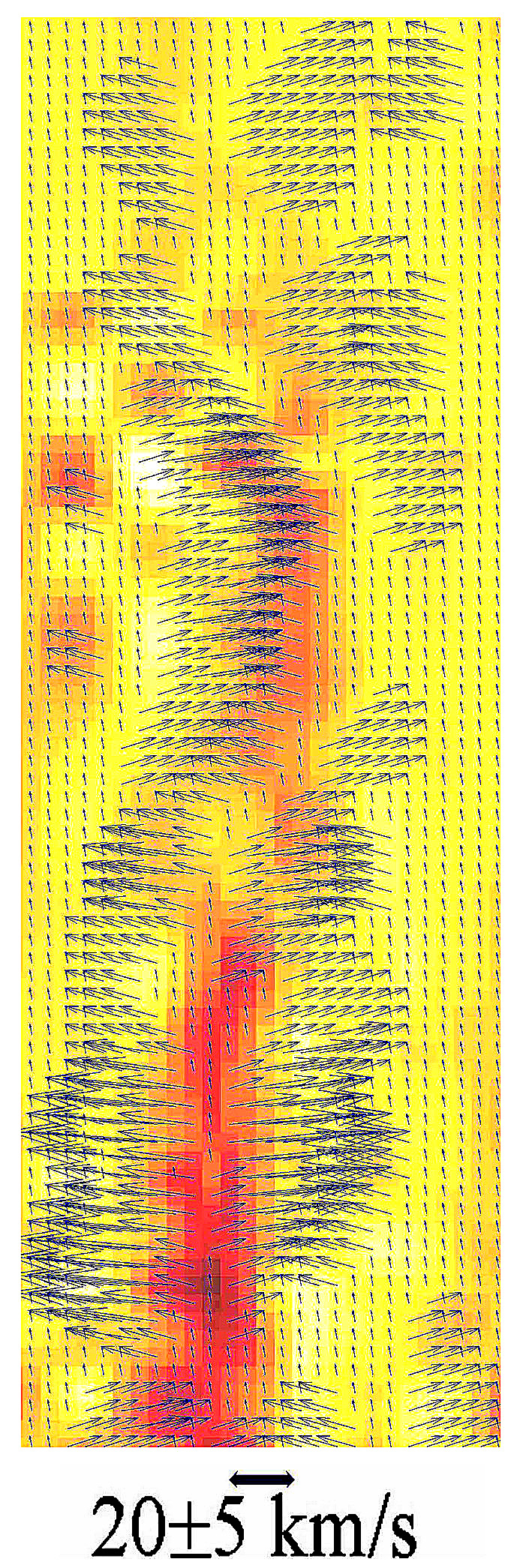}
      \caption{Zoom of a small part of the preceding Fig.~\ref{figure3}, bottom of the last frame, showing at the bottom the scale of apparent velocities measured using the FLCT algorithm at ultimate resolution.  Pixel size is 0.1 arcsec and time cadence only 1.6 sec. Note the artifact produced outside the structures with arrows pointing upward. The method is not suitable for measuring a proper motion when the velocity is aligned along an extended fine structure like the case of a vertical, up or down, flow along a quasi-radial spicule.
              }
         \label{figure5}
\end{figure}

    It works with two images separated in time and it gives an estimate for a 2D velocity field by finding the shifts that increase a local cross-correlation function between the two images. Pixel size was 0.1 arcsec and matrices were shifted by the same value to compute crosss-correlations. Using the FLCT algorithm for each spicule-case we then deduce a two-dimensional velocity diagram as shown in Fig.~\ref{figure3},~\ref{figure4} and~\ref{figure5}. After evaluating all maps, we find that a large percentage of solar coronal hole spicules shows i/ a surge-like behavior (see Section 5 for a description of "surge"), and ii/ support an interpretation as twisting multi-component spicules. The processed accompanying movie corresponding to the June 17, 2009 observations (6 min 45 sec long with a 3 sec cadence) brings another support to our results.\\


\begin{figure}
   \centering
   \includegraphics[width=\hsize]{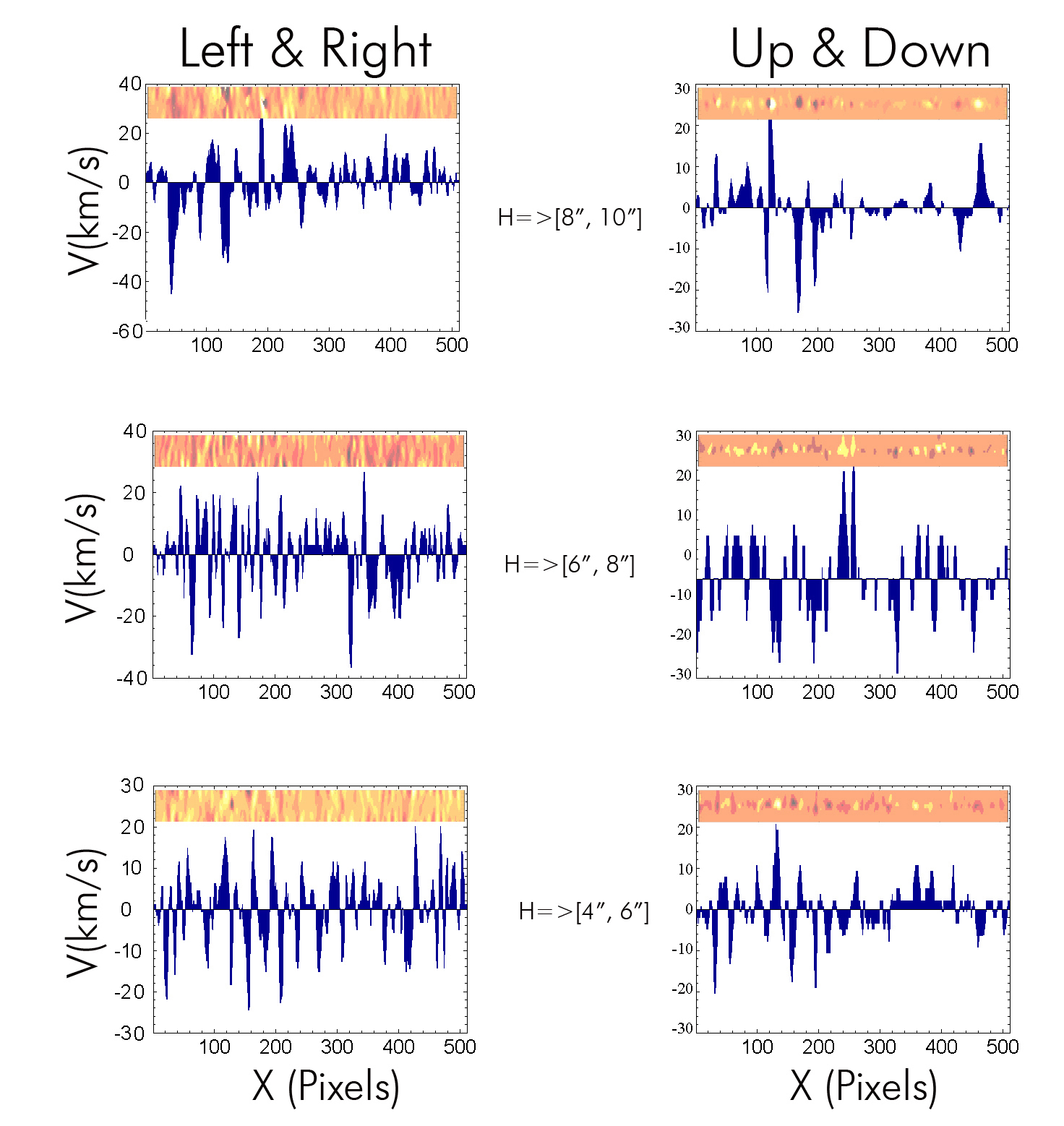}
      \caption{Distribution of horizontal and vertical proper motion “velocities” for different layers or slices above the limb as given by the height “H” in arcsec above the surface (limb) as observed on 17 June 2011. The orange band at top shows the area selected at different heights and the plots show the dispersion around the mean values .
              }
         \label{figure6}
\end{figure}

 We detected several long spicules showing a similar behavior and found that the upward and downward flows are equal for lower and middle levels, see Fig.~\ref{figure6}, but the rate of upward motion is slightly larger in high levels. However, we cannot really trust the deduced values of up and down velocities because the FLCT algorithm used at the ultimate resolution is poorly responding to velocities directed along a fine structure which is predominantly directed along the local radial direction (vertical). As suggested by Fig.~\ref{figure2}, large amplitude velocities could exist. As far as the shearing motion in left and right directions is concerned, it is also equal at all heights.\\
  The features are delicate to measure precisely because the signal/noise ratio is rather marginal and the variations are extremely fast. The frames are separated by only 3 sec and the field covers just 5$\times$5 Mm$^{2}$. The neo-spicule feature (shown with a white inclined bar / ) at 05:17:00 is rapidly rising vertically until 05:17:06 UT and then seems to slowly fall during at least the 6 following sec.  The second neo-spicule feature (shown with a white  >) at 05:17:30 rises rapidly with a small angle to the vertical until 05:17:36 and fades after and then reinforces at 05:17: 51 with a small evidence of twist 6 sec later at 05:17:57 UT. The apparent associated upwards velocities are typically 250 $\pm$ 100 kms$^{-1} $in both cases, taking into account the scale (given in Mm) and the pixel size of 0.1 arcsec or 74 km in July.  Note the bright low lying feature under the 694 Mm height that could be a small loop. The horizontal signature of an artifact of the CCD reading at the 694.7 Mm and the 696.1 Mm heights appears as a noise difficult to remove.\\
  In Fig.~\ref{figure6}, orange strips velocity diagram put at the top are obtained using the FLCT algorithm, used in vertical (up and down) and horizontal (left and right) direction and the blue curves under each strip give the corresponding velocity values. Up and down velocities are probably under-estimated because structures are predominantly aligned in the vertical direction.\\

\begin{figure}
   \centering
   \includegraphics[width=\hsize]{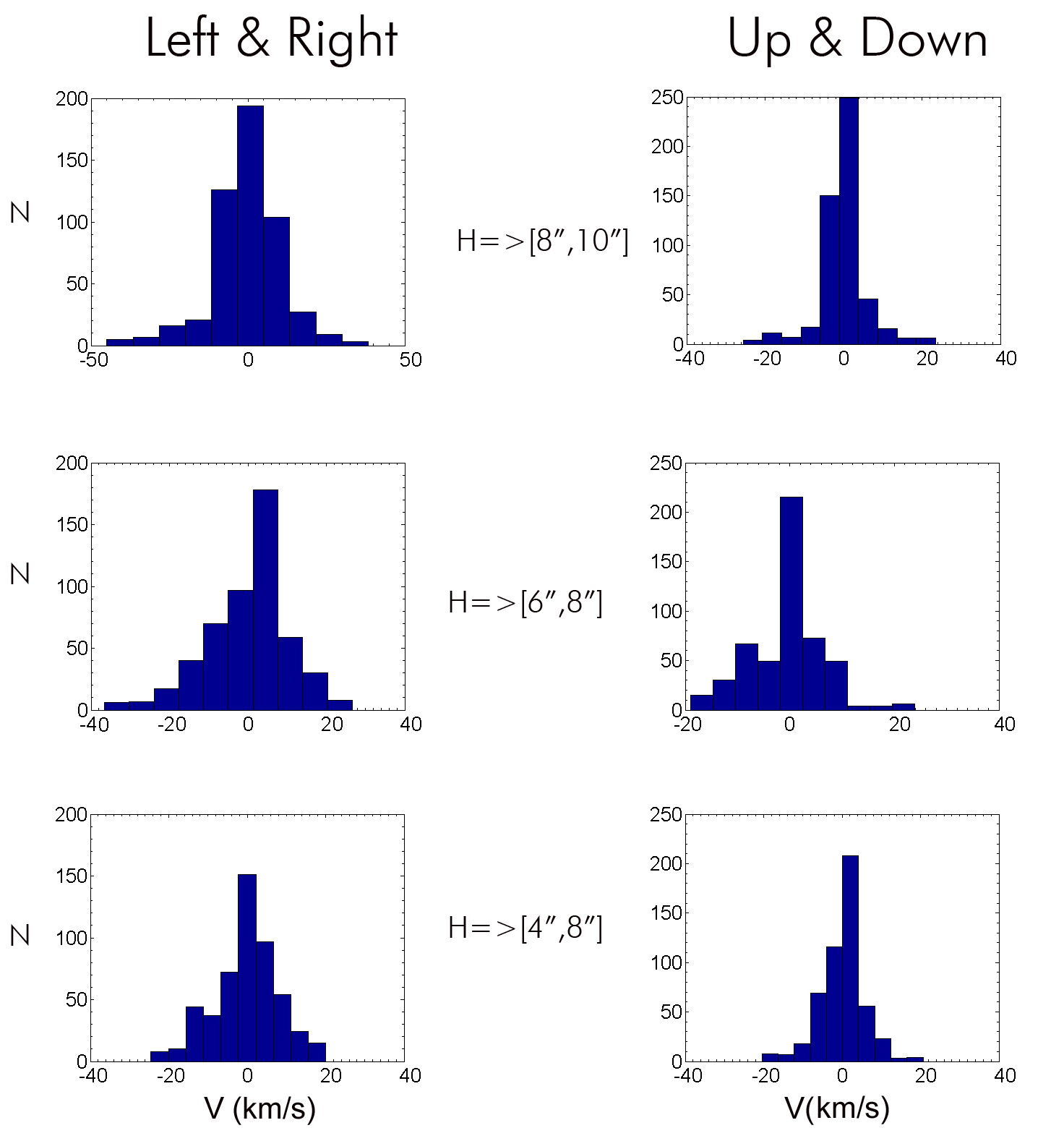}
      \caption{Histograms for swaying and up and down motions for different heights observed on 17 June 2011.
              }
         \label{figure7}
\end{figure}

   Finally, we plotted histograms of velocities. We found the amplitudes are increasing with the height (Fig.~\ref{figure7}) for the left and right-hand velocities.\\

   The analysis is repeated for another sequence with different images and data taken on 25 Oct. 2008.  \\

\begin{figure}
   \centering
   \includegraphics[width=\hsize]{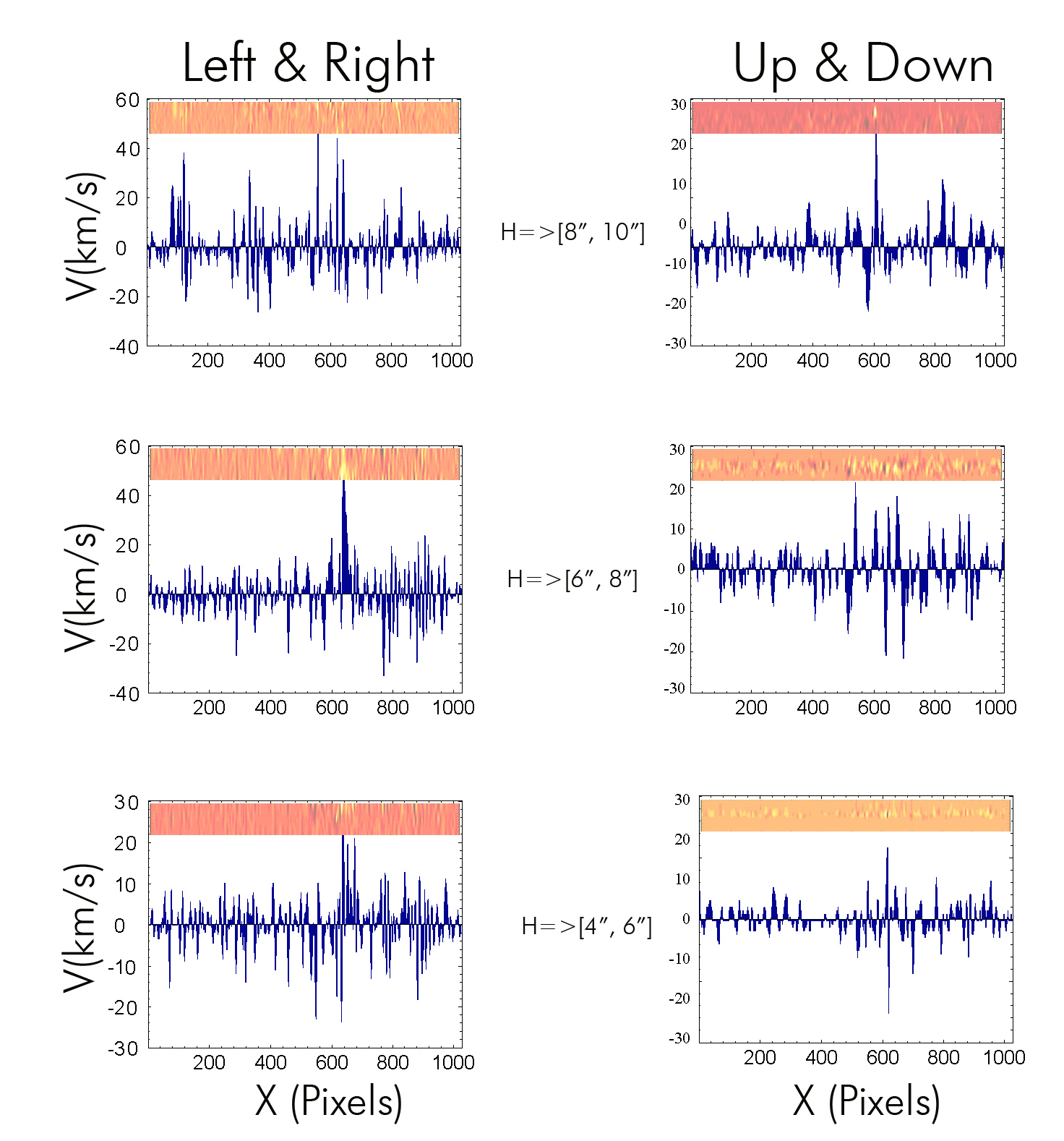}
      \caption{Horizontal and vertical proper motion velocities for different layers above the limb as observed on 25 Oct. 2008.
              }
         \label{figure8}
\end{figure}

\begin{figure}
   \centering
   \includegraphics[width=\hsize]{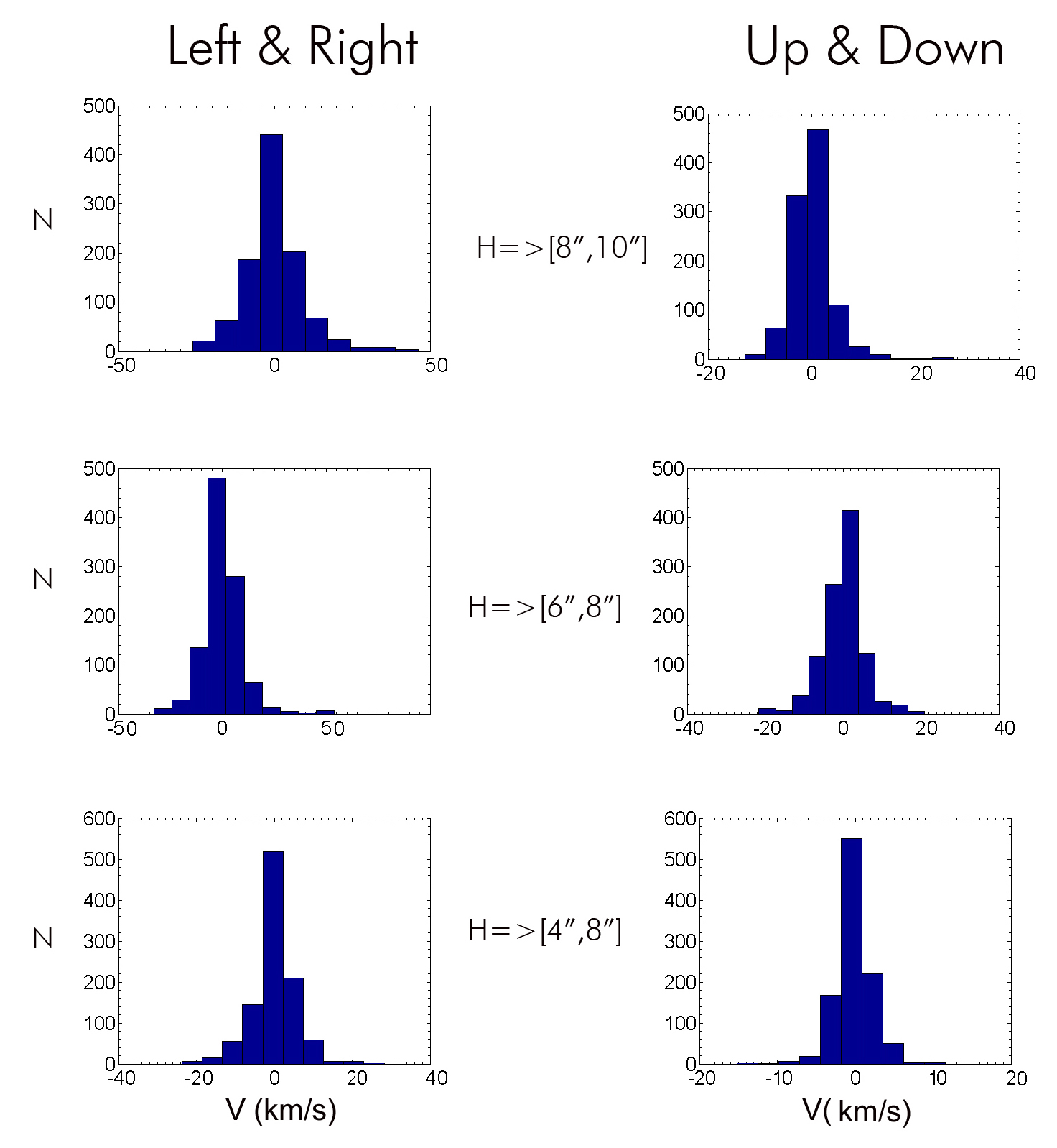}
      \caption{Histograms for swaying and up and down motions for different heights on 25 Oct. 2008.
              }
         \label{figure9}
\end{figure}%

\section{Discussion}
\begin{figure}
   \centering
   \includegraphics[width=5cm]{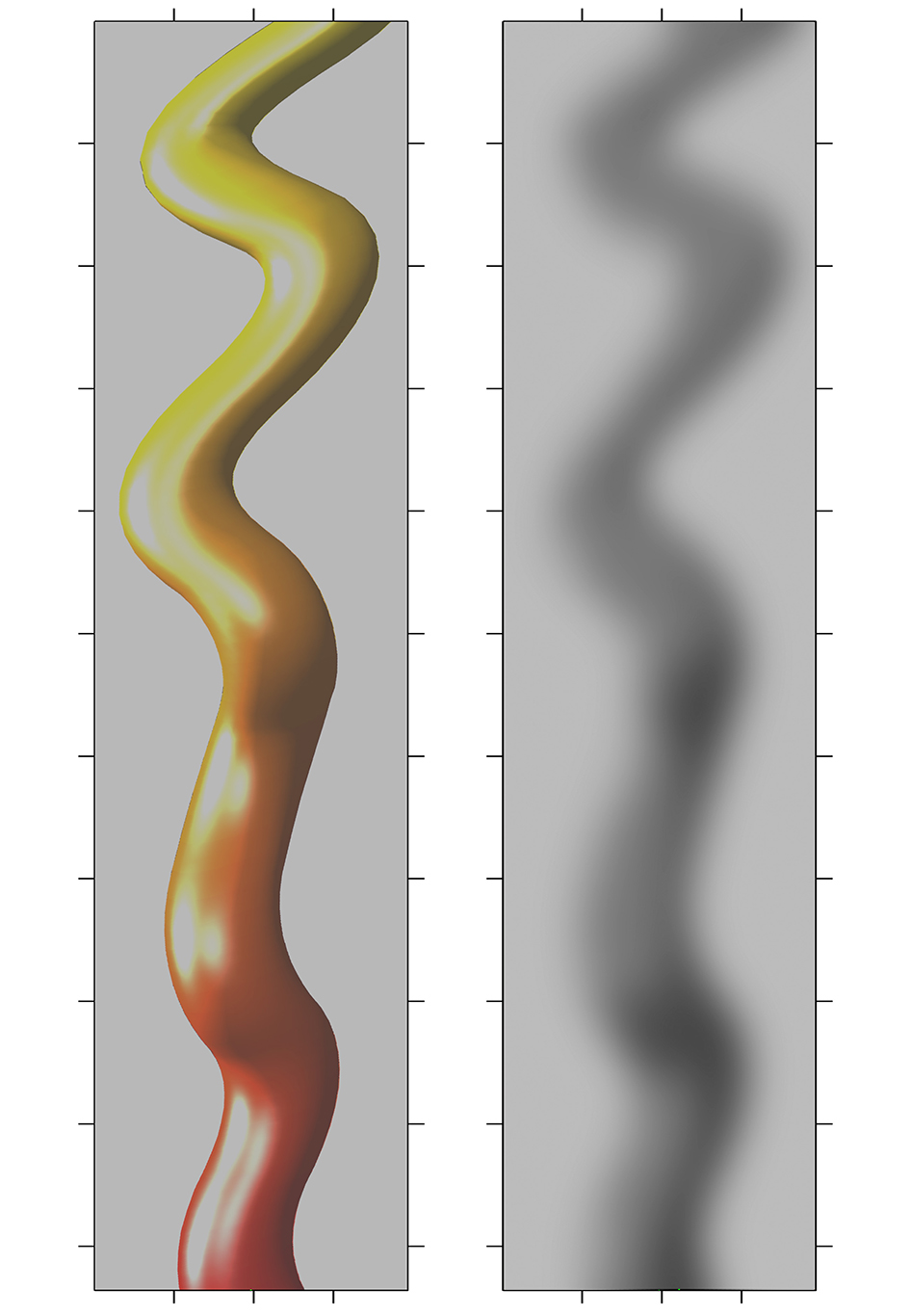}
      \caption{Cartoon showing a twisted thread (at left) using a 3-D effect with propagation of torsional motion, and in the right panel, the corresponding view in 2-D assuming the plasma is optically thin at this emission line (shown in negative). }
                       \label{figure10}
\end{figure}

    Generally speaking, a large number of multicomponent spicules is observed. It comes immediately in mind that it is possibly comparable to surge-like events observed in polar regions that are also called macro-spicules (e.g. \citealp{Georgakilasb}). Because torsional motion occurs simultaneously and close to each another components (Fig.~\ref{figure6} to~\ref{figure9}), it might be interpreted as twisting threads, see Fig.~\ref{figure10} for a naive representation).  According to the 2-D image of spicule, it is thought that the waves are Kink mode, but according to the 3-D display, wave propagation in a tube is seen as Helical and Kink, what we call Helical-Kink because the spicule axis is displaced. The number of twists depends on the diameter of the set of components with a coherent behavior and changes from at least 1 turn for very thin structures to at most 2 or 3 turns for mini-surge-like very broad structure. A curvature shape looking similar to a transverse kink mode oscillation along the threads is possibly produced by a low twist number.\\

  Fig.~\ref{figure1} shows an example of what could be long rotating and untwisting Macro-spicules. We suggest that such spicules have a multi-component structure and are created due to the disruption of a primary loop (\citealp{Tavabib}). In the polar regions spicules are longer and less tilted (\citealp{Tavabid}), so the probability of observing individual spicules increases. Also, the lower density of the background in the polar regions helps us to better see spicule foot-points.\\

  Surges are cool plasma jets ejected from small flare-like chromospheric bright points, including regions near the poles, such as sub-flares. Some similarity possibly exists with Ellerman bombs (moustaches) that usually appear near sunspots at lower latitudes. These cool plasma jets are a kind of active small prominence, usually observed in H$\alpha$ at ground-based observatories (\citealp{Georgakilasa} and \citealp{Georgakilasb}) and known for a long time from eclipse observations (e.g. Fig. 140, p. 389 in \citealp{Secchi}), although space observations, such as with EUV telescopes, also detect surges. Surges in TR lines are ejected at peak velocities of 50–200 kms$^{-1}$ along a straight or slightly curved path, delineating magnetic field lines. They reach heights of 10000–100000 km and typically last 10–20 min. Some of the surges show spinning motion at a few tens of kms$^{-1}$ around their axis, during the ejection phase, see the classical work of \cite{Rompolt} for an exhaustive description of H$\alpha$ coronagraphic observations. The direction of spinning motion is consistent with the direction of unwinding motion of their helical twist. Small surges are then similar to twisty spicules, which are cool chromospheric jets observed in H$\alpha$ and EUV at the network boundary. The velocity and height are about 30 to 250 kms$^{-1}$ and 5000–8000 km at most, much smaller than those of surges, while the temperature and electron density are comparable with those of surges. Macro-spicules also show a velocity feature suggesting spinning motion (\citealp{Georgakilasa} and \citealp{Georgakilasb}). Macrospicules are observed in coronal holes and show characteristic parameters between those of spicules and those of surges. They are ejected at velocities of order of 140 kms$^{-1}$, reaching heights of 10000– 35000 km, and show nearly ballistic motion (\citealp{Tavabib} and \citealp{Tsiropoula}). Spinning motion can also be interpreted taking into account reconnections between the twisted flux tube and the untwisted magnetic field of the background corona. Therefore it is very likely that magnetic reconnection is a key to producing surges as it has also been suggested for coronal polar jet (\citealp{Patsourakos}). If a magnetic field line reconnects with another field line, a magnetic tension force is generated in the reconnected field line as in a slingshot (described in e.g. \citealp{Tavabic}). Such a magnetic tension force accelerates plasmas in the reconnected field line up to the Alfvén speed outside the current sheet.\\

  We note that in case jet-spicules are cool enough, it is difficult to accelerate such cool jets by gas pressure alone without raising the temperature, whereas it is quite easy to accelerate cool jets by magnetic force while keeping a low temperature. Hence, the magnetically driven jet mechanism would have a wider applicability as a model for both cool jets and for hot X-Ray jets (\citealp{Filippovc}). This is especially true for larger jets (such as large surges and sprays), although, as for smaller jets (such as spicules and small surges), it remains a possibility that jets are accelerated by the gas pressure force of a slow-mode MHD shock.\\

  The study of spinning spicules seems important for the following reasons:\\
  a) Because of conservation of the torsional magnetic fields, it can give us information about the origin of spicules.\\
  b) Our work presented here and other studies made using both filtergrams and limb spectra (\citealp{Pasachoff}; \citealp{Nikolsky}; \citealp{Tanaka}; \citealp{Kulidzanishvili}; \citealp{Gadzhiev}; \citealp{Suematsua}; \citealp{Dara}; \citealp{Georgakilasa} and recently in \citealp{Tavabia}) clearly suggests that spicules are multi-component. At each moment, we usually can only see clearly just a part of it; accordingly the spatial and the temporal resolution should be better than the values of the diameter of each component and its lifetime. We should emphasis that the spinning motions of limb spicules have been already reported in the literature (\citealp{Pasachoff}; \citealp{Georgakilasb} and  \citealp{Tavabi}) with the suggestion, from time sequences of photographic limb spectra taken with the giant soviet era 52 cm aperture Lyot coronagraph, that the spinning trajectory is elliptical (\citealp{Gadzhiev}). Spinning motion in erupting prominences is more easily studied (\citealp{Rompolt}). \\
  c) Propagation of Alfvénic waves in spicules and especially further out along the magnetic field can be considered for heating the corona and for supplying the material for the solar wind.\\

  \citet*{Antolin} reported the behavior of multi-stranded loops with a coherent evolution throughout their lifetime in the chromosphere; several recent reports confirm the multi-strand loop scenario using the AIA and especially the High-resolution Coronal Imager (Hi-C), where an  unprecedented spatial resolution of about of 150 km was obtained in coronal line emission (\citealp{Brooks} and \citealp{Peter}) with the evidence of  braiding and series of reconnection sites along the highly twisted loops (\citealp{Cirtain}).\\

  To understand the origin of the twist of the loops, several scenarios were presented; each of them attempts to explain the origin of this behavior using different assumptions. It seems that untwist results from a tearing of initially helically structured loops (Fig.~\ref{figure1}). In these images a loop was seen at the feet of twisted spicules. Based on the conservation of rotational momentum (or helicities), we can expect that the expansion of chromospheric mini-loops finally leads to micro-eruptions similar to spicule eruption of the spine type (\citealp{Tavabic}). To go further with observations, some better signal/noise high resolution frames are needed (Fig.~\ref{figure2}), which means larger aperture telescopic observations as now available with the NST at BBSO and those planned with the ATST at Hawaii and with the future ISAS Solar C space telescope.  Meanwhile, numerical simulations would help a lot and we look forward to evaluating more simulations of the spicule phenomenon, starting from the surface of the Sun where  granular and sub-granular motions and possibly p-mode interactions in the presence of a background magnetic field of the chromopsheric network would produce helical small scale eruptions  (\citealp{Kitiashvili}) and inclusion of dissipation effects resulting from the local dynamo effects in converging flows (\citealp{Lorraina} and \citealp{Lorrainb}) will be taken into account.\\

\begin{acknowledgements}
We are grateful to the Hinode team for their wonderful observations. Hinode is a Japanese mission developed and launched by ISAS/JAXA, with NAOJ as domestic partner and NASA, ESA and STFC (UK) as international partners. Image processing software was provided by O. Koutchmy see http://www.ann.jussieu.fr/$\sim$koutchmy/index\_newE.html. The FLCT source code and compilation instructions were downloaded from http://solarmuri.ssl.berkeley.edu/overview/publicdownloads/software.html.\\
This work has been supported by Research Institute for Astronomy \& Astrophysics of Maragha (RIAAM) and the Center for International Scientific Studies \& Collaboration (CISSC), by the French Embassy in Tehran and the French Institut d’Astrophysique de Paris-CNRS and UPMC. We thank Leon Golub for his careful reading of the draft and for meaningful suggestion and Alphonse Sterling for comments and suggestions. Last but not least, we thank the Referee(s) and the Editor for meaningful suggestions and requests that greatly improved the original draft.
\end{acknowledgements}


\end{document}